\documentclass[a4paper]{jpconf}
\usepackage{graphicx}
\begin{document}

\title{%
Charge-spin-orbital dynamics of one-dimensional two-orbital Hubbard model}

\author{Hiroaki Onishi}

\address{%
Advanced Science Research Center,
Japan Atomic Energy Agency,
Tokai, Ibaraki 319-1195, Japan}


\begin{abstract}

We study the real-time evolution of a charge-excited state
in a one-dimensional $e_{\rm g}$-orbital degenerate Hubbard model,
by a time-dependent density-matrix renormalization group method.
Considering a chain along the $z$ direction,
electrons hop between adjacent $3z^2$$-$$r^2$ orbitals,
while $x^2$$-$$y^2$ orbitals are localized.
For the charge-excited state,
a holon-doublon pair is introduced into the ground state
at quarter filling.
At initial time,
there is no electron in a holon site,
while a pair of electrons occupies $3z^2$$-$$r^2$ orbital in a doublon site.
As the time evolves,
the holon motion is governed by the nearest-neighbor hopping,
but the electron pair can transfer
between $3z^2$$-$$r^2$ orbital and $x^2$$-$$y^2$ orbital
through the pair hopping
in addition to the nearest-neighbor hopping.
Thus holon and doublon propagate at different speed
due to the pair hopping that is characteristic of multi-orbital systems.

\end{abstract}

\section{Introduction}

Competing interaction
among charge, spin, and orbital degrees of freedom
plays a crucial role in the emergence of various types of quantum phases
in strongly correlated electron systems
with active orbital degree of freedom
\cite{Tokura2000,Dagotto2001,Hotta2006}.
The change of conditions,
such as temperature, magnetic field, pressure, and chemical doping,
gives rise to a dramatic change of
transport and magnetic properties through a phase transition
because of a subtle balance between multiple phases.

Recent developments in femto-second laser technology have made it possible
to investigate the dynamical control of the many-body electron state
on an ultrafast time scale,
the so-called photo-induced phase transition
\cite{Tokura2006}.
In general,
the photo-irradiation can excite the system to a novel non-equilibrium state
that cannot be accessed by simply changing temperature,
since photon energy is much higher than thermal energy.
A photo-excited non-equilibrium state eventually relaxes
to the original equilibrium state.
However, the system exhibits a transient behavior
and sometimes it decays into a metastable state
rather than the original state.
In Pr$_{0.7}$Ca$_{0.3}$MnO$_{3}$,
a charge/orbital-ordered insulating state is melted
and a ferromagnetic metallic state is induced by the photo-irradiation
\cite{Miyaho1997,Fiebig1998}.
Such a photo-induced insulator-to-metal transition
has also been demonstrated
in Nd$_{0.5}$Ca$_{0.5}$MnO$_{3}$
\cite{Ogasawara2002}.

In this paper,
to gain deep insight into the ultrafast non-equilibrium dynamics of
the complex charge-spin-orbital state,
we investigate the real-time dynamics of a charge-excited state
in an $e_{\rm g}$-orbital model
from the viewpoint of the time evolution of wavepackets
by exploiting numerical techniques.
We discuss the effects of the pair-hopping interaction
on the propagation of holon and doublon.

\section{Model and numerical method}

Let us consider an $e_{\rm g}$-orbital degenerate Hubbard model
on the linear chain along the $z$ direction
with $N$ sites and one electron per site (quarter filling),
described by
\begin{eqnarray}
\label{eq:H}
 H
 &=&
 \sum_{i,\tau,\tau',\sigma}
 t_{\tau\tau'}
 (d_{i\tau\sigma}^{\dag} d_{i+1\tau'\sigma}+\mbox{h.c.})
 +U \sum_{i,\tau}
 \rho_{i\tau\uparrow} \rho_{i\tau\downarrow}
 +U' \sum_{i,\sigma,\sigma'}
 \rho_{i\alpha\sigma} \rho_{i\beta\sigma'}
 \nonumber\\
 &&
 +J \sum_{i,\sigma,\sigma'}
 d_{i\alpha\sigma}^{\dag} d_{i\beta\sigma'}^{\dag}
 d_{i\alpha\sigma'} d_{i\beta\sigma}
 +J' \sum_{i,\tau\neq\tau'}
 d_{i\tau\uparrow}^{\dag} d_{i\tau\downarrow}^{\dag}
 d_{i\tau'\downarrow} d_{i\tau'\uparrow},
\end{eqnarray}
where $d_{i\tau\sigma}$
is an annihilation operator for an electron 
with spin $\sigma$ (=$\uparrow,\downarrow$)
in orbital $\tau$ (=$\alpha$ for $3z^2$$-$$r^2$; $\beta$ for $x^2$$-$$y^2$)
at site $i$,
and $\rho_{i\tau\sigma}$=%
$d_{i\tau\sigma}^{\dag}d_{i\tau\sigma}$.
The hopping amplitude is given by
$t_{\alpha\alpha}$=$1$ (taken as energy unit),
$t_{\alpha\beta}$=$t_{\beta\alpha}$=$t_{\beta\beta}$=0.
Note that due to the orbital anisotropy,
$3z^2$$-$$r^2$ orbital becomes itinerant,
while $x^2$$-$$y^2$ orbital is localized.
Regarding the on-site Coulomb interaction,
$U$ is the intra-orbital Coulomb interaction,
$U'$ the inter-orbital Coulomb interaction,
$J$ the inter-orbital exchange interaction,
and $J'$ the pair-hopping interaction between different orbitals.
The relation $U$=$U'$+$J$+$J'$ holds
due to the rotation symmetry in the local orbital space
and $J$=$J'$ is assumed due to the reality of the orbital function
\cite{Dagotto2001},
so that we have two independent interaction parameters.
In this paper,
we set $U'$=$10$ and study the dependence on $J'$.
We use the unit such that $\hbar$=$1$,
and the time is measured in units of $\hbar/t_{\alpha\alpha}$.

We investigate the dynamics of the charge-excited state
from the viewpoint of the propagation of wavepackets
\cite{Onishi2009}.
For this purpose, we exploit
density-matrix renormalization group (DMRG) techniques.
First, the ground state $\vert\psi_{0}\rangle$
is obtained by an ordinary static DMRG method
\cite{White1992}.
The finite-system algorithm is employed
with the use of open boundary conditions.
Then, as an initial state at time $t$=$0$,
we prepare a charge-excited state by creating a holon-doublon pair,
defined as
\begin{equation}
 \vert\psi_{1}^{hd}\rangle =
 \sum_{\tau\tau'\sigma}
 t_{\tau\tau'} c_{i\tau\sigma}^{\dag} c_{i+1\tau'\sigma} \,
 \vert\psi_{0}\rangle.
\end{equation}
The time evolution of $\vert\psi_{1}^{hd}\rangle$ is computed
by an adaptive time-dependent DMRG method
\cite{Daley2004,White2004}.
The second-order Suzuki-Trotter decomposition is used
to describe the time-evolution operator as the product of
local time-evolution operators
with a small time step $\Delta t$=$0.02$.
We keep up to $m$=$300$ states in renormalization steps,
and the truncation error is kept below $10^{-7}$
during the time evolution.

\section{Results}

\begin{figure}[t]
\begin{center}
\includegraphics[width=0.9\textwidth]{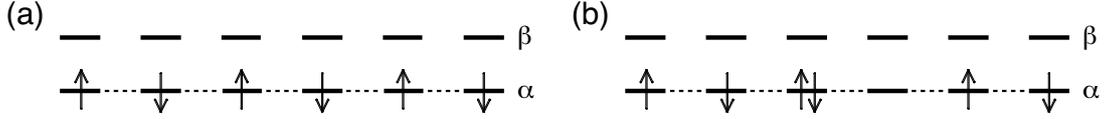}
\end{center}
\caption{
Electron configuration of
(a) ground state and
(b) charge-excited state.
Dotted lines indicate hopping connection
between adjacent $\alpha$ orbitals
}
\end{figure}

In Fig.~1(a),
we depict the electron configuration of the ground state.
Since the electron hopping connects $\alpha$ orbitals
in adjacent sites,
$\alpha$ orbital is singly occupied in every site
so as to gain kinetic energy.
Localized $S$=$1/2$ spins exhibit
quasi-long-range antiferromagnetic order
due to the antiferromagnetic exchange interaction.
For the charge-excited state,
a holon-doublon pair is created at the center of the chain,
as shown in Fig.~1(b).
We note that $\alpha$ orbital is doubly occupied in a doublon site.

\begin{figure}[t]
\begin{center}
\includegraphics[width=0.9\textwidth]{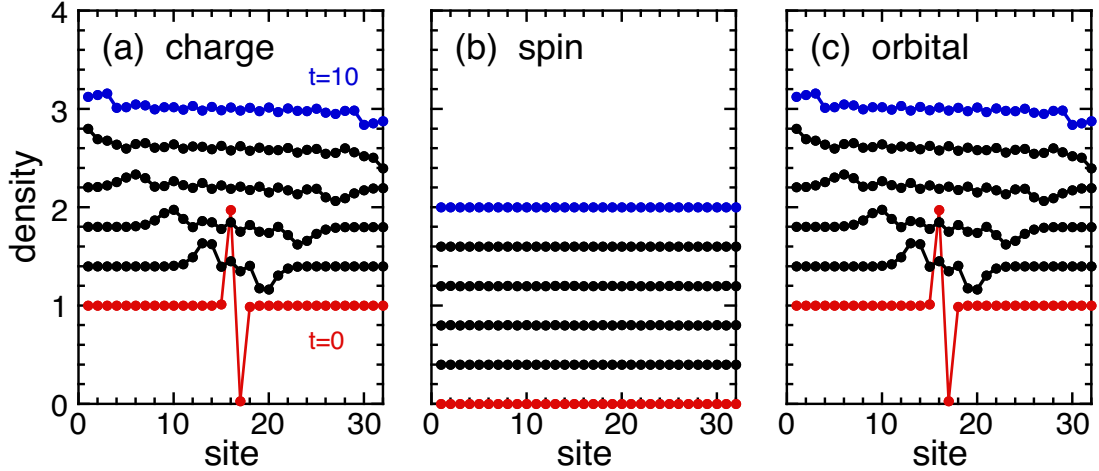}
\end{center}
\caption{
Time evolution of
(a) charge,
(b) spin, and
(c) orbital densities
at $U'$=$10$ and $J'$=$0$.
A holon-doublon pair is introduced
at the center of the chain of $N$=$32$.
Time-series results from time $t$=$0$ to $t$=$10$ are arranged
with shifted upward along the vertical axis.
}
\end{figure}

To gain insight into the dynamics of the charge-excited state,
we measure the time evolution of
the charge, spin, and orbital densities, defined by
$n_{c}(i,t)$=$\sum_{\tau\sigma} \langle \rho_{i\tau\sigma} \rangle_{t}$,
$n_{s}(i,t)$=$\sum_{\tau} \langle \rho_{i\tau\uparrow}$$-$$\rho_{i\tau\downarrow} \rangle_{t}$,
and
$n_{o}(i,t)$=$\sum_{\sigma} \langle \rho_{i\alpha\sigma}$$-$$\rho_{i\beta\sigma} \rangle_{t}$,
where $\langle \cdots \rangle_{t}$ denotes the expectation value
using the wavefunction at time $t$.
In Fig.~2, we present our numerical results at $J'$=$0$.
Concerning the charge density,
a holon-doublon pair is completely localized at the chain center at $t$=$0$,
i.e., $n_{c}(N/2,0)$$\simeq$$2$ and $n_{c}(N/2$+$1,0)$$\simeq$$0$,
as shown in Fig.~2(a).
As the time evolves, holon and doublon wavepackets expand
in the system.
We see that the wavefront of the holon wavepacket propagates toward right,
while that of the doublon wavepacket moves left,
but there is no fine structure in the region between two wavefronts
due to interference effects.
Note that holon and doublon wavepackets propagate
with the same speed,
since holon and doublon move
through the electron hopping of the same amplitude $t_{\alpha\alpha}$.
On the other hand, as shown in Fig.~2(b),
the spin density is not affected by the charge excitation,
since we assume $S_{tot}^{z}$=$0$ for the charge-excited state,
where $S_{tot}^{z}$ is the $z$ component of total spin.
As shown in Fig.~2(c),
the orbital density is equivalent to the charge density.
Once two electrons occupy $\alpha$ orbital in the doublon site,
$\beta$ orbital is always vacant
due to the absence of the pair-hopping process at $J'$=$0$,
so that
$n_{c}(i,t)$=$n_{o}(i,t)$=$\sum_{\sigma}\langle\rho_{i\alpha\sigma}\rangle_{t}$.

At finite $J'$, however,
a pair of electrons accommodated in $\alpha$ orbital
can transfer to $\beta$ orbital through the pair hopping,
leading to a difference in the propagation of holon and doublon.
In fact, we find that the doublon wavepacket propagates
at half speed in comparison to the holon wavepacket,
as shown in Fig.~3(a).
Let us here discuss the effects of the pair hopping
on the doublon propagation.
Assuming that the electron pair occupies
either $\alpha$ orbital or $\beta$ orbital,
the local two-electron state is described
by the linear combination of
$\vert i\alpha$$\uparrow\rangle$$\vert i\alpha$$\downarrow\rangle$
and
$\vert i\beta$$\uparrow\rangle$$\vert i\beta$$\downarrow\rangle$,
where
$\vert i\tau\sigma\rangle$=$d_{i\tau\sigma}^{\dag}\vert 0\rangle$
and $\vert 0\rangle$ is the vacuum state.
With these two states, the eigenstates are given by
$\vert ia\rangle$$\equiv$%
$(\vert i\alpha$$\uparrow\rangle$$\vert i\alpha$$\downarrow\rangle$%
$-$$\vert i\beta$$\uparrow\rangle$$\vert i\beta$$\downarrow\rangle)/\sqrt{2}$
and
$\vert ib\rangle$$\equiv$%
$(\vert i\alpha$$\uparrow\rangle$$\vert i\alpha$$\downarrow\rangle$%
$+$$\vert i\beta$$\uparrow\rangle$$\vert i\beta$$\downarrow\rangle)/\sqrt{2}$
with its eigenenergy
$E_{a}$=$U$$-$$J'$ and $E_{b}$=$U$+$J'$, respectively.
Among them, we suppose that
the lowest-energy state $\vert ia\rangle$
represents the doublon state in the strong-coupling limit.
Then, doublon and single electron can be exchanged
by an effective hopping,
\begin{equation}
 \langle i+1 \alpha\sigma' \vert
 \langle ia \vert
 t_{\alpha\alpha}
 (c_{i\alpha\sigma}^{\dag}c_{i+1\alpha\sigma}+{\rm h.c.})
 \vert i\alpha\sigma' \rangle
 \vert i+1 a \rangle
 = -\frac{1}{2}t_{\alpha\alpha}\delta_{\sigma,-\sigma'}.
\end{equation}
Namely, the hopping amplitude for doublon is reduced
to half of the original hopping amplitude
due to the pair hopping.
On the other hand, holon and single electron are exchanged
with the original hopping amplitude.
Thus doublon propagates at half speed comparing with holon.

For the spin state,
no structure appears in the spin density even at $J'$=$2$,
as shown in Fig.~3(b).
Figure.~3(c) presents the orbital density.
We can see a correspondence of the orbital density
to the charge density,
while they are not equivalent any longer.
In fact, regarding the right-moving wavefront,
the orbital density is quite similar to the charge density.
The left-moving wavefront of the orbital density
propagates with the same speed as that of
the left-moving wavefront of the charge density.
However,
the difference of the orbital density from the uniform value
varies between positive and negative
as the time evolves,
indicating that the electron pair actually occupies
not only $\alpha$ orbital but also $\beta$ orbital.
Thus the orbital structure around doublon is dynamically deformed
due to the pair hopping.

\begin{figure}[t]
\begin{center}
\includegraphics[width=0.9\textwidth]{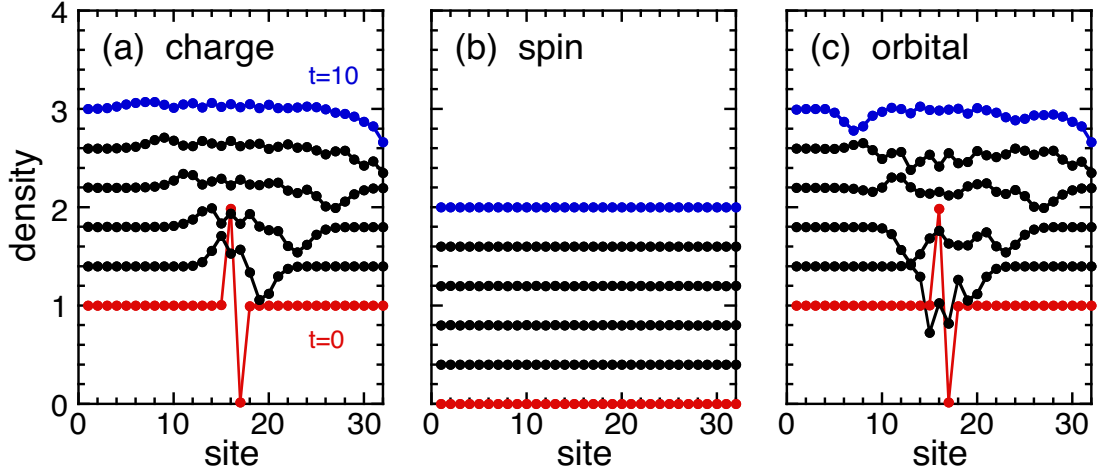}
\end{center}
\caption{
Time evolution of
(a) charge,
(b) spin, and
(c) orbital densities
at $U'$=$10$ and $J'$=$2$.
}
\end{figure}

\section{Summary}

We have studied the time evolution of the charge-excited state
of the one-dimensional $e_{\rm g}$-orbital degenerate Hubbard model
by using DMRG techniques.
We have shown that
doublon propagates at half speed in comparison to the holon motion
because of the pair-hopping process
peculiar to multi-orbital systems.

\ack

The author would like to thank T. Fujii and K. Ueda for discussions.


\end{document}